\documentclass[aps,amssymb,amsfonts,amsmath]{revtex4}
\usepackage{times}
\usepackage{revsymb}
\usepackage{bm}
\usepackage{mathptm}
\usepackage[english]{babel}
\usepackage{mathrsfs}
\usepackage{hyperref}
\hypersetup{pdftitle={On Extremal Quantum States of Composite Systems with Fixed Marginals},
  pdfsubject={x},
  pdfauthor={Oliver Rudolph
  },
  pdfkeywords={acrobat},
  colorlinks={true}
  }

\begin{document}
\small
\normalsize
\newcounter{saveeqn}
\newcommand{\alpheqn}{\setcounter{saveeqn}{\value{equation}}%
\setcounter{equation}{0}%
\renewcommand{\theequation}{\mbox{\arabic{saveeqn}-\alph{equation}}}}
\newcommand{\reseteqn}{\setcounter{equation}{\value{saveeqn}}
\renewcommand{\theequation}{\arabic{equation}}}
\setcounter{equation}{1}
\protect\newtheorem{theo}{Theorem}
\protect\newtheorem{prop}{Proposition}
\protect\newtheorem{lem}{Lemma}
\protect\newtheorem{co}{Corollary}
\newtheorem{cri}{Criterion}
\protect\newtheorem{de}{Definition}
\protect\newtheorem{con}{Conjecture}
\newtheorem{ex}{Example}
\newtheorem{rema}{Remark}
\newtheorem{rem}{Remark}
\small
\normalsize
\title{On extremal quantum states of composite systems with fixed marginals}
\author{Oliver Rudolph}
\affiliation{Quantum Information Theory Group,  Dipartimento di Fisica ``A.~Volta,'' \\
Universit\`a degli Studi
di Pavia, via Bassi 6, I-27100 Pavia, Italy}
\begin{abstract}
\noindent We study the convex set $\mathcal{C}(\varrho_{1}, \varrho_{2})$ of all bipartite quantum states with fixed marginal states $\varrho_{1}$ and $\varrho_{2}$. The extremal states in this set have recently been characterized by Parthasarathy [Ann.~Henri Poincar\'e (to appear), quant-ph/0307182, \cite{Parthasarathy03}]. Here we present an alternative necessary and sufficient condition for a state in $\mathcal{C}(\varrho_{1}, \varrho_{2})$ to be extremal. Our approach is based on a canonical duality between bipartite states and a certain class of completely positive maps and has the advantage that it is easier to check and to construct explicit examples of extremal states. In dimension $2 \times 2$ we give a simple new proof for the fact that all extremal states in $\mathcal{C}\left(\frac{1}{2} \openone, \frac{1}{2} \openone \right)$ are precisely the projectors onto maximally entangled wave functions. We also prove that in higher dimension this does not hold and construct an explicit example of an extremal state in $\mathcal{C}\left(\frac{1}{3} \openone, \frac{1}{3} \openone \right)$ that is not maximally entangled. Generalizations of this result to higher dimensions are also discussed.
 \end{abstract}
\maketitle
\texttt{To appear in  \href{http://jmp.aip.org}{Journal of Mathematical Physics}}
\section{Introduction}
In the paradigmatic situation encountered in quantum information processing two or more (often spatially separated) parties share the different parts of a composite quantum system. The parties are able to perform arbitrary operations on their respective parts ``locally'' and to communicate classically among each other to orchestrate their actions. The fundamental realization in quantum information theory is that sharing the parts of a composite quantum system can enable the parties to perform certain communication or information processing tasks more efficiently than classically (see \cite{NielsenC} for an introduction). 
Mathematically this setting raises a number of new and interesting structural questions. Among them the study of quantum channels and the characterization of quantum entanglement play a central role \cite{Werner,Bruss02,DonaldHR02}. The present letter is devoted to the characterization of the set of quantum states with fixed marginal states. This problem was recently posed and studied in  detail by Parthasarathy \cite{Parthasarathy03}. Let $\mathscr{H}_{1}$ and $\mathscr{H}_{2}$ be two finite dimensional complex Hilbert spaces, corresponding to two finite level quantum systems $S_{1}$ and $S_{2}$. Without loss of generality we assume that $d := \dim(\mathscr{H}_{1}) = \dim(\mathscr{H}_{2}).$ [Otherwise we embed the lower dimensional Hilbert spaces into the larger one.] The states for $S_{i}$ are given by the positive operators on $\mathscr{H}_{i}$ with trace one. We denote the set of all states on $\mathscr{H}_{i}$ by $\mathtt{S}(\mathscr{H}_{i}).$ The composite quantum system $S_{12}$ of $S_{1}$ and $S_{2}$ is described by the tensor product $\mathscr{H}_{1} \otimes \mathscr{H}_{2}$.  A \emph{state} for $S_{12}$ is a positive operator on $\mathscr{H}_{1} \otimes \mathscr{H}_{2}$ with trace one. The space of all states is denoted by $\mathtt{S}(\mathscr{H}_{1} \otimes \mathscr{H}_{2}).$ Consider $\varrho \in \mathtt{S}(\mathscr{H}_{1} \otimes \mathscr{H}_{2})$. The \emph{reductions} or \emph{marginal states} of $\varrho$ are given by $\varrho_{1} := \mathrm{tr}_{2}(\varrho) \in 
\mathtt{S}(\mathscr{H}_{1})$ and $\varrho_{2} := \mathrm{tr}_{1}(\varrho) \in \mathtt{S}(\mathscr{H}_{2})$. Here $\mathrm{tr}_{1}$ and $\mathrm{tr}_{2}$ denote the partial traces over $\mathscr{H}_{1}$ and $\mathscr{H}_{2}$ respectively.
Now fix $\varrho_{1} \in \mathtt{S}(\mathscr{H}_{1})$ and $\varrho_{2} \in \mathtt{S}(\mathscr{H}_{2})$. We denote by $\mathcal{C}(\varrho_{1}, \varrho_{2})$ the convex set of all states $\varrho \in \mathtt{S}(\mathscr{H}_{1} \otimes \mathscr{H}_{2})$ whose marginal states are equal to $\varrho_{1}$ and $\varrho_{2}$ respectively. The set of extreme points of $\mathcal{C}(\varrho_{1}, \varrho_{2})$ will be denoted by $\mathcal{E}(\varrho_{1}, \varrho_{2}).$ 
Throughout this paper we will denote the set of all operators on a Hilbert space $\mathscr{H}$ by $\mathtt{L}(\mathscr{H})$. The identity in $\mathtt{L}(\mathscr{H})$ is denoted by $\openone$, or, when $\mathscr{H}$ is $d$-dimensional, by $\openone_{d}$. Slightly abusing the notation we will also denote the identity map from $\mathtt{L}(\mathscr{H})$ into itself by $\openone$.

In his work \cite{Parthasarathy03} Parthasarathy presented a necessary and sufficient condition for an element $\varrho \in \mathcal{C}(\varrho_{1}, \varrho_{2})$ to be an extreme point. This was then used to derive an upper bound on the rank of such an extremal state. In the special case $\mathscr{H}_{1} = \mathscr{H}_{2} = \mathbb{C}^2$ and $\varrho_{1} = \varrho_{2} = \frac{1}{2} \openone_{2}$, Parthasarathy found that a state $\varrho \in \mathcal{C}\left( \frac{1}{2} \openone_{2},\frac{1}{2} \openone_{2} \right)$ is extremal if and only if it is a projector onto the subspace spanned by a maximally entangled wavefunction. 
A wavefunction in $\mathbb{C}^2 \otimes \mathbb{C}^2$ is called maximally entangled if it is of the form $\vert \psi_{+} \rangle = \frac{1}{\sqrt{2}}\left( \vert 0 \rangle \vert \phi_{0} \rangle + \vert 1 \rangle \vert \phi_{1} \rangle \right)$ where $\{ \vert 0 \rangle, \vert 1 \rangle \}$ denotes the canonical basis of $\mathbb{C}^2$ and where $\{ \vert \phi_{0} \rangle, \vert \phi_{1} \rangle\}$ is any other orthonormal basis of $\mathbb{C}^2$.
For higher dimensions the question of whether or not there are extremal states with maximally mixed marginals  -- i.e., states in $\mathcal{E}\left(\frac{1}{d} \openone, \frac{1}{d}\openone \right)$ -- that are not projectors onto maximally entangled wavefunctions was left open in \cite{Parthasarathy03}.

In the present letter we present an alternative approach to the characterization of $\mathcal{E}(\varrho_{1}, \varrho_{2})$ that transforms the problem into that of finding the extreme points of a certain convex set of completely positive maps that satisfy an additional requirement. This will allow us to derive an alternative necessary and sufficient condition for a state $\varrho \in \mathcal{C}(\varrho_{1}, \varrho_{2})$ to be extremal. We will then study the special case of states with maximally mixed marginals, i.e., when $\varrho_{1} = \varrho_{2} = \frac{1}{d} \openone$. For $d=2$ we will give a simple proof for Parthasarathys result that the extremal states are exactly the projectors onto maximally entangled wavefunctions. For $d>2$ our results imply that there are extremal states in $\mathcal{E}\left(\frac{\openone}{d}, \frac{\openone}{d}\right)$ that are not projectors onto maximally entangled pure states. We give an explicit example for an extremal state on $\mathbb{C}^3 \otimes \mathbb{C}^3$ with maximally mixed marginals that is not equal to a projector onto a maximally entangled wavefunction. Finally we we discuss generalizations of this result to higher dimensions.

\section{Duality between bipartite states and completely positive maps} \label{sec2}
The approach in the present paper relies upon a duality between bipartite quantum states on $\mathscr{H}_{1} \otimes \mathscr{H}_{2}$ and completely positive maps $\Lambda : \mathtt{L}(\mathscr{H}_{2}) \to \mathtt{L}(\mathscr{H}_{1})$ that preserve the trace of the completely mixed state, i.e., that satisfy $\mathrm{tr}\left(\Lambda\left(\frac{1}{d} \openone\right)\right) = 1$ (this is very often called the \emph{Jamio\l{}kowski isomorphism}, see \cite{Jamiolkowski72} 
and for a related duality \cite{Werner}). A map $\Lambda : \mathtt{L}(\mathscr{H}_{2}) \to \mathtt{L}(\mathscr{H}_{1})$ is called completely positive if $\Lambda \otimes \openone  : \mathtt{L}(\mathscr{H}_{2} \otimes \mathscr{K}) \to \mathtt{L}(\mathscr{H}_{1} \otimes \mathscr{K})$ is positive for any finite dimensional ancilla Hilbert space  $\mathscr{K}$.

We make the identification $\mathscr{H}_{1} \simeq \mathbb{C}^{d}$ and $\mathscr{H}_{2} \simeq \mathbb{C}^{d}$. In other words, we pick orthonormal bases in $\mathscr{H}_{1}$ and $\mathscr{H}_{2}$ and identify them with the canonical real basis in $\mathbb{C}^{d}$ and $\mathbb{C}^{d}$ respectively.  We denote these bases by $\{ \vert i \rangle_{1} \}_{i=1}^{d}$ and $\{\vert i \rangle_{2} \}_{i=1}^{d}$ respectively. Finally we introduce the maximally entangled pure wavefunction 
\[ \vert \psi_{+} \rangle := \frac{1}{\sqrt{d}} \sum_{i=1}^d \vert i \rangle_{2} \vert i \rangle_{2} \in \mathscr{H}_{2} \otimes 
\mathscr{H}_{2}. \]

The duality between bipartite state and completely positive maps depends explicitly on this choice for the canonical bases. Let $\Lambda : \mathtt{L}(\mathscr{H}_{2}) \to \mathtt{L}(\mathscr{H}_{1})$ be a completely positive map with $\mathrm{tr}\left(\Lambda\left(\frac{1}{d} \openone\right)\right) = 1.$ Then \alpheqn \begin{equation} \label{e1a}
\varrho_{\Lambda} := \Lambda \otimes \openone(\vert \psi_{+} \rangle \langle \psi_{+} \vert)
\end{equation} defines a bipartite state on $\mathscr{H}_{1} \otimes \mathscr{H}_{2}$. The complete positivity of $\Lambda$ ensures that $\varrho \geq 0$ while the condition $\mathrm{tr}(\Lambda(\frac{1}{d} \openone)) = 1$ ensures that $\mathrm{tr}(\varrho_{\Lambda}) =1$.

Conversely, let $\varrho$ be a bipartite state on $\mathscr{H}_{1} \otimes \mathscr{H}_{2}$. Then
\begin{equation} \label{e1b} \Lambda_{\varrho} (\sigma) := d \mathrm{tr}_{2} [(\openone \otimes \sigma^{\mathtt{T}} \varrho)]
\end{equation} \reseteqn defines a completely positive map $\Lambda_{\varrho} : \mathtt{L}(\mathscr{H}_{2}) \to \mathtt{L}(\mathscr{H}_{1})$ that satisfies $\mathrm{tr}\left(\Lambda_{\varrho}\left(\frac{1}{d} \openone\right)\right) = 1$. Here $^\mathtt{T}$ denotes the transposition with respect to the canonical real basis. By explicit calculation one checks that for a given $\Lambda$ we have $\Lambda_{\varrho_{\Lambda}} = \Lambda$ and for a given $\varrho$ we have $\varrho_{\Lambda_{\varrho}} = \varrho$. Thus the correspondence $\Lambda \leftrightarrow \varrho$ described by Equations (\ref{e1a}) and (\ref{e1b}) is bijective \cite{Jamiolkowski72}. 

\section{Joint linear independence} \label{sec3}
To formulate the main result in this paper it is useful to introduce the concept of \emph{joint linear independence} of two families of vectors. In the following definition $X^{\times r}$ denotes the $r$-fold cartesian product of the set $X$ by itself.
\begin{de} 
Let $V$ and $W$ be complex vector spaces. Then two ordered $r$-tuples $(v_{i})_{i=1}^r \in V^{\times r}$ and $(w_{i})_{i=1}^r \in W^{\times r}$ are called \emph{jointly linearly independent} if the family $\{ v_{i} \oplus w_{i} \}_{i=1}^r$ in the direct sum $V \oplus W$ is a linearly independent family.  
\end{de}
Notice that this definition depends on the order of the $r$-tuples. 
The following is an immediate consequence of the definition.
\begin{lem} \label{l2} Let $V$ and $W$ be complex vector spaces and  let $(v_{i})_{i=1}^r \in V^{\times r}$ and $(w_{i})_{i=1}^r \in W^{\times r}$ be two ordered $r$-tuples of vectors. If $\{ v_{i} \}_{i=1}^r$ is linearly independent in $V$ or if $\{ w_{i} \}_{i=1}^r$ is linearly independent in $W$, then $(v_{i})_{i=1}^r$ and $(w_{i})_{i=1}^r$ are jointly linearly independent.
\end{lem}
Notice that the converse implication does not hold in general. If $\{ v_{i} \}_{i=1}^r$ is linearly dependent in $V$ and if $\{ w_{i} \}_{i=1}^r$ is linearly dependent in $W$, then $\{ v_{i} \oplus w_{i} \}_{i=1}^r$ is not necessarily linearly dependent in $V \oplus W$.
\begin{lem} \label{l3}
Let $V$ be a complex *-algebra and let $(v_{j})_{j=1}^r \in V^{\times r}$ be an ordered $r$-tuple of elements. If $\{ v_{j} \}_{j}$ is linearly dependent, then the $r^2$-tuples $(v_{i}^* v_{j})_{ij}$ and $(v_{j} v_{i}^*)_{ij}$ cannot be jointly linearly independent.
\end{lem}
\emph{Proof}. Since $\{ v_{j} \}_{j}$ is linearly dependent, there exist $(\lambda_{j})_{j} \in \mathbb{C}^r$ such that $\lambda_{j_{0}} \neq 0$ for some $j_{0}$ and $\sum_{j=1}^r \lambda_{j} v_{j} = 0$. Therefore also
$\sum_{ij} \delta_{i i_{0}} \lambda_{j} (v_{i}^* v_{j}, v_{j} v_{i}^*) =0$ for all $i_{0}$. $\Box$ 

\section{Extremal states in $\mathcal{C}(\varrho_{1}, \varrho_{2})$}
Let $\varrho \in \mathcal{C}(\varrho_{1}, \varrho_{2})$. In $\mathscr{H}_{2}$ consider an orthonormal basis of Eigenvectors of $\varrho_{2}$, i.e., $\varrho_{2} = \sum_{i} r_{i} \vert r_{i} \rangle \langle r_{i} \vert$. We  identify the basis $\{ \vert r_{i} \rangle \}_{i=1}^d$ of Eigenvectors of $\varrho_{2}$ with the canonical real basis of $\mathscr{H}_{2} \simeq \mathbb{C}^{d}$. Further we write
\begin{equation} \label{maxent} \vert \psi_{+} \rangle := \frac{1}{\sqrt{d}} \sum_{i} \vert r_{i} \rangle \otimes \vert r_{i} \rangle.  \end{equation}
In the sequel it is always understood that the bijection between states and completely positive maps from Section \ref{sec2} is with respect to this choice of the canonical basis and that the maximally entangled state in Equation (\ref{e1a}) is the state from Eq.~(\ref{maxent}). To every state $\varrho \in \mathcal{C}(\varrho_{1}, \varrho_{2})$ Eq.~(\ref{e1b}) gives a unique completely positive map $\Lambda_{\varrho}$ that satisfies \addtocounter{equation}{1} \alpheqn
\begin{eqnarray} \label{c1a} \Lambda_{\varrho}(\openone) & = & d \varrho_{1}, \\
\label{c2} \Lambda'_{\varrho}(\openone) & = & d \varrho_{2}. 
\end{eqnarray} \reseteqn Here $\Lambda'_{\varrho}$ denotes the canonical dualization of $\Lambda_{\varrho}$ defined by
$\mathrm{tr}(\Lambda'_{\varrho}(x)y) = \mathrm{tr}(x \Lambda_{\varrho}(y))$ for all $y$.
In terms of the Kraus representation of $\Lambda_{\varrho}(x) = \sum_{j} V_{j}^\dagger x V_{j}$ the conditions (\ref{c1a}) and (\ref{c2}) can be expressed as  \addtocounter{equation}{1} \alpheqn
\begin{eqnarray} \sum_{j} V_{j}^\dagger V_{j} & = & d \varrho_{1}, \\
\sum_{j} V_{j} V_{j}^\dagger & = & d \varrho_{2}. 
\end{eqnarray} \reseteqn
We denote the set of all completely positive maps $\Lambda : \mathtt{L}(\mathscr{H}_{2}) \to \mathtt{L}(\mathscr{H}_{1})$ satisfying the conditions (\ref{c1a}) and (\ref{c2}) by $\mathtt{CP}(\mathscr{H}_{2}, \mathscr{H}_{1}, \varrho_{1}, \varrho_{2})$. It is clear that $\mathtt{CP}(\mathscr{H}_{2}, \mathscr{H}_{1}, \varrho_{1}, \varrho_{2})$ is a convex set. The bijection described in Eqs.~(\ref{e1a}) and (\ref{e1b}) obviously respects the convex structure. In particular it establishes a bijection between $\mathcal{E}(\varrho_{1},\varrho_{2})$ 
 and the extreme point of $\mathtt{CP}(\mathscr{H}_{2}, \mathscr{H}_{1}, \varrho_{1}, \varrho_{2})$. 

We are now ready to state our main result
\begin{theo} \label{t1} Let $\Lambda : \mathtt{L}(\mathscr{H}_{2}) \to \mathtt{L}(\mathscr{H}_{1})$ be a completely positive map in $\mathtt{CP}(\mathscr{H}_{2}, \mathscr{H}_{1}, \varrho_{1}, \varrho_{2})$. Then $\Lambda$ is extreme in $\mathtt{CP}(\mathscr{H}_{2}, \mathscr{H}_{1}, \varrho_{1}, \varrho_{2})$ if and only if  $\Lambda$ admits an expression $\Lambda(x) = \sum_{j} V_{j}^\dagger x V_{j}$ for all $x \in \mathtt{L}(\mathscr{H}_{2})$, where $V_{i}$ are $d \times d$ matrices, satisfying the following conditions
\begin{itemize} \item $\sum_{j} V_{j}^\dagger V_{j} = d \varrho_{1}$, \item $\sum_{j} V_{j} V_{j}^\dagger = d \varrho_{2}$, \item  $(V_{i}^\dagger V_{j})_{ij}$ and $(V_{j} V_{i}^\dagger)_{ij}$ are jointly linearly independent.
\end{itemize}
\end{theo}
For the proof of Theorem \ref{t1} we need the following lemma. For a proof see Remark 4 in \cite{Choi75}.
\begin{lem} \label{lemmachoi}
Let $\Lambda$ be a completely positive map with Kraus representation $\Lambda(x)=\sum_{j} V_{j}^\dagger x V_{j}$ with $\{ V_{j} \}^\ell_j$ linearly independent. Let $\{ W_{p} \}^{\ell'}_{p}$ be a class of $ d \times d$ matrices, then $\Lambda$ has the expression $\Lambda(x) = \sum_{p}^{\ell'} W_{p}^\dagger x W_{p}$ if and only if there exists an isometric $\ell'  \times \ell$ matrix $(\mu_{pi})_{pi},$ such that $W_{p} = \sum_{i} \mu_{pi} V_{i}$ for all $p.$
\end{lem}
\emph{Proof of Theorem \ref{t1}}. The proof is an only slight modification and generalization of the proof of Theorem 5 in \cite{Choi75}. We include it for the convenience of the reader. First assume that $\Lambda$ is extremal in $\mathtt{CP}(\mathscr{H}_{2}, \mathscr{H}_{1}, \varrho_{1}, \varrho_{2})$. We express $\Lambda$ in Kraus form $\Lambda(x) = \sum_{j} V_{j}^\dagger x V_{j}$. Without loss of generality we can assume that $\{ V_{j} \}_{j}$ is linearly independent \cite{Choi75}. Now suppose that $\sum \lambda_{ij} V_{i}^\dagger V_{j} = 0$ and $\sum_{ij} \lambda_{ij} V_{j} V_{i}^\dagger =0.$ We need to show that $\lambda_{ij} =0.$  Without loss of generality we can assume that $(\lambda_{ij})_{ij}$ is a hermitean matrix and $- \openone \leq (\lambda_{ij})_{ij} \leq \openone$ (for details see \cite{Choi75}). 

Define $\Phi _{\pm} : \mathtt{L}(\mathscr{H}_{2}) \to \mathtt{L}(\mathscr{H}_{1})$  by
$\Phi_{\pm}(x) := \sum_{j} V_{j}^\dagger x V_{j} \pm \sum_{ij} \lambda_{ij} V_{i}^\dagger x V_{j}.$ Hence $\Phi_{\pm}(\openone) = d \varrho_{1}$ and $\Phi'_{\pm}(\openone) = d \varrho_{2}.$ We set $\openone + (\lambda_{ij})_{ij} 
= (\alpha_{ij})^\dagger_{ij} (\alpha_{ij})_{ij} \geq 0$ and $W_{i} := \sum_{j} \alpha_{ij} V_{j}.$ By direct computation, $\Phi_{+}(x) = 
\sum_{i} W_{i}^\dagger x W_{i}$. Hence $\Phi_{+}$ is completely positive. Similarly it can be shown that $\Phi_{-}$ is completely positive. Since $\Lambda$ is extremal, we find that $\Lambda = \Phi_{+}$. Therefore by Lemma \ref{lemmachoi} $(\alpha_{ij})_{ij}$ is an isometry and $\openone + (\lambda_{ij})_{ij} = \openone$. This implies $(\lambda_{ij})_{ij} =0.$

Now assume that $\Lambda$ admits a representation of the form $\Lambda(x) = \sum_{j} V_{j}^\dagger x V_{j}$ for all $x \in \mathtt{L}(\mathscr{H}_{2})$ where $\sum_{j} V_{j}^\dagger V_{j} = d \varrho_{1}$, $\sum_{j} V_{j} V_{j}^\dagger = d \varrho_{2}$, and  $(V_{i}^\dagger V_{j})_{ij}$ and $(V_{j} V_{i}^\dagger)_{ij}$ are jointly linearly independent. By Lemma \ref{l3} also $\{ V_{j} \}_{j}$ is linearly independent. Now suppose $\Lambda = \frac{1}{2} (\Phi_{1} + \Phi_{2})$ with $\Phi_{1}(x)= \sum_{p} W_{p}^\dagger x W_{p}$, $\Phi_{2}(x)=\sum_{q} Z_{q}^\dagger x Z_{q}$, and
$\sum_{p} W_{p}^\dagger W_{p} = \sum_{q} Z_{q}^\dagger Z_{q} = d \varrho_{1}$,  $\sum_{p} W_{p} W_{p}^\dagger = \sum_{q} Z_{q} Z_{q}^\dagger = d \varrho_{2}.$ Since $\Lambda(x)= \frac{1}{2} \sum_{p} W_{p}^\dagger x W_{p} + \frac{1}{2} \sum_{q} Z_{q}^\dagger x Z_{q},$ it follows by Lemma \ref{lemmachoi} that $W_{p}$ and $Z_{q}$ can be expressed as a linear combination of the $V_{j}$. Let $W_{p} = \sum_{i} \mu_{pi}V_{i}$ for all $p$. Then $\sum_{j} V_{j}^\dagger V_{j} = \sum_{p} W_{p}^\dagger W_{p} = \sum_{pij} \mu_{pi}^* \mu_{pj} V_{i}^\dagger V_{j}$ and $\sum_{j} V_{j} V_{j}^\dagger = \sum_{p} W_{p} W_{p}^\dagger = \sum_{pij} \mu_{pi}^* \mu_{pj} V_{j} V_{i}^\dagger.$ The joint linear independence of 
$(V_{i}^\dagger V_{j})_{ij}$ and $(V_{j} V_{i}^\dagger)_{ij}$ implies $\sum_{p} \mu_{pi}^* \mu_{pj} = \delta_{ij}.$ In other words $(\mu_{pi})_{pi}$ is an isometry. By Lemma \ref{lemmachoi}, we conclude that $\Lambda = \Phi_{1}.$ Thus $\Lambda$ is extremal in  $\mathtt{CP}(\mathscr{H}_{2}, \mathscr{H}_{1}, \varrho_{1}, \varrho_{2})$. $\square$

\begin{co} \label{c1}
Let $\varrho \in \mathcal{C}(\varrho_{1}, \varrho_{2}).$ Write the spectral decomposition of $\varrho_{2}$ as $\varrho_{2} = \sum_{i} r_{i} \vert r_{i} \rangle \langle r_{i} \vert.$ Then $\varrho \in \mathcal{E}(\varrho_{1}, \varrho_{2})$ if and only if there exists a family of $d \times d$ matrices $\{V_{j} \}$ such that $\varrho$ can be expressed as
\[ \varrho = \frac{1}{d} \sum_{i j k} V_{j}^\dagger \vert r_{i} \rangle \langle r_{k} \vert V_{j} \otimes \vert r_{i} \rangle \langle r_{k} \vert \] where $\{V_{j} \}_{j}$ satisfy the following conditions
\begin{itemize} \item $\sum_{j} V_{j}^\dagger V_{j} = d \varrho_{1}$, \item $\sum_{j} V_{j} V_{j}^\dagger = d \varrho_{2}$, \item  $(V_{i}^\dagger V_{j})_{ij}$ and $(V_{j} V_{i}^\dagger)_{ij}$ are jointly linearly independent.
\end{itemize}
\end{co}

\begin{rem} \label{r6} Suppose $\Lambda : \mathtt{L}(\mathscr{H}_{2}) \to \mathtt{L}(\mathscr{H}_{1})$ is completely positive. Then we can write $\Lambda(x) = \sum_{j} V_{j}^\dagger  x V_{j}$ where $\{ V_{j} \}_{i=1}^\ell$ is a class of linearly independent $d \times d$ matrices. Therefore $\ell \leq d^2$. If $\Lambda$ is extremal in $\mathtt{CP}(\mathscr{H}_{2}, \mathscr{H}_{1}, \varrho_{1}, \varrho_{2})$ we can conclude that $\ell \leq \sqrt{2} d$. Indeed, $(V_{i}^\dagger V_{j})_{ij}$ and $(V_{j} V_{i}^\dagger)_{ij}$ are jointly linearly independent only if the cardinal number of $\{ V_{i}^\dagger V_{j} \oplus V_{j} V_{i}^\dagger \}_{ij}$ is smaller than $\dim(\mathtt{L}(\mathscr{H}_{2})) + \dim(\mathtt{L}(\mathscr{H}_{1}))$. In other words $\ell^2 \leq 2 d^2$, i.e., $\ell \leq \sqrt{2} d$. Parthasarathy found a slightly stronger bound in \cite{Parthasarathy03}: $\ell \leq \sqrt{2 d^2 -1}.$ It is not known whether this bound is tight.
\end{rem}

\begin{rem} The bound $\ell \leq \sqrt{2} d$ also implies that for any $\varrho \in \mathcal{E}(\varrho_{1}, \varrho_{2})$ we have $\mathtt{rank}(\varrho) \leq \sqrt{2} d$. In all dimensions $d \geq 2$ this implies that any $\varrho \in \mathcal{E}(\varrho_{1}, \varrho_{2})$ is singular.
\end{rem}

\section{Examples}
\subsection{A two dimensional example}
Consider $\mathbb{C}^2 \otimes \mathbb{C}^2$ and the convex set $\mathcal{C}\left(\frac{1}{2} \openone, \frac{1}{2} \openone \right)$ of states on $\mathbb{C}^2 \otimes \mathbb{C}^2$ with maximally mixed marginals. This is a physically interesting example. It was previously studied in \cite{Parthasarathy03}.

Assume that $\varrho \in \mathcal{E}\left(\frac{1}{2} \openone, \frac{1}{2} \openone \right)$, i.e., that $\varrho$ is extremal in $\mathcal{C}\left(\frac{1}{2} \openone, \frac{1}{2} \openone \right)$. By Corollary \ref{c1} there is a linearly independent family of $2 \times 2$ matrices $\{ V_{i} \}^\ell_{i=1}$ such that 
\[ \varrho = \frac{1}{2} \sum_{i j k} V_{j}^\dagger \vert r_{i} \rangle \langle r_{k} \vert V_{j} \otimes \vert r_{i} \rangle \langle r_{k} \vert \] where $\{V_{j} \}$ satisfy the following conditions
\addtocounter{equation}{1} \alpheqn
\begin{eqnarray} \label{e4a} \sum_{j} V_{j}^\dagger V_{j} & = & \openone_{2}, \\
\label{e4b} \sum_{j} V_{j} V_{j}^\dagger & = & \openone_{2}, \\ \nonumber
 \end{eqnarray} \reseteqn
and where $(V_{i}^\dagger V_{j})_{ij}$ and $(V_{j} V_{i}^\dagger)_{ij}$ are jointly linearly independent. 

By Remark \ref{r6} either $\ell = 1$ or $\ell = 2$. In the case $\ell = 1$, the matrix $V_{1}$ is unitary and it follows from Corollary \ref{c1} that $\varrho$ is equal to the projector onto the subspace spanned by a maximally entangled wavefunction.

Now consider the case $\ell =2$. Consider the singular value decompositions of $V_{1}$ and $V_{2}$ respectively, i.e., $V_{1} = \sum_{s=1}^2 \sqrt{v_{s}(1)} \vert \varphi_{s} \rangle \langle \psi_{s} \vert$ and $V_{2} = \sum_{s=1}^2 \sqrt{v_{s}(2)} \vert \varphi'_{s} \rangle \langle \psi'_{s} \vert$, where $v_{s}(i)$ are non-negative coefficients and where $\{ \vert \psi_{s} \rangle \}_{s=1}^2, \{ \vert \psi'_{s} \rangle \}_{s=1}^2, \{ \vert \varphi_{s} \rangle \}_{s=1}^2$ and $\{ \vert \varphi'_{s} \rangle \}_{s=1}^2$ are four orthonormal bases of $\mathbb{C}^2$.  Then $V_{1}^\dagger V_{1} = \sum_{s=1}^2 v_{s}(1) \vert \psi_{s} \rangle \langle \psi_{s} \vert$, $V_{2}^\dagger V_{2} = \sum_{s=1}^2 v_{s}(2) \vert \psi'_{s} \rangle \langle \psi'_{s} \vert$, $V_{1} V_{1}^\dagger = \sum_{s=1}^2 v_{s}(1) \vert \varphi_{s} \rangle \langle \varphi_{s} \vert$ and $V_{2} V_{2}^\dagger = \sum_{s=1}^2 v_{s}(2) \vert \varphi'_{s} \rangle \langle \varphi'_{s} \vert$. 

First consider the case of degenerate singular values, i.e., assume $v_{1}(1) = v_{2}(1)$. Then $V_{1}^\dagger V_{1} = V_{1} V_{1}^\dagger = v_{1}(1) \openone$ and $V_{2}^\dagger V_{2} = V_{2} V_{2}^\dagger = v_{1}(2) \openone$. Moreover, Equations (\ref{e4a}) and (\ref{e4b}) imply that $v_{1}(1) = 1 - v_{1}(2).$ However this implies that $(V_{i}^\dagger V_{j})_{ij}$ and $(V_{j} V_{i}^\dagger)_{ij}$ are not jointly linearly independent. By Corollary \ref{c1} $\varrho$ is not extremal in $\mathcal{C}\left(\frac{1}{2} \openone, \frac{1}{2} \openone \right)$. This is a contradiction.

Secondly, consider the case of non-degenerate singular values, i.e., $v_{1}(1) \neq v_{2}(1)$. In this case Equations (\ref{e4a}) and (\ref{e4b}) imply that $v_{s}(1) = 1 -v_{s}(2)$, $\vert \varphi_{s} \rangle = \vert \varphi'_{s} \rangle$ and $\vert \psi_{s} \rangle = \vert \psi'_{s} \rangle$ for $s=1,2$. By direct computation it is easily verified that $V_{1}^\dagger V_{2} = V_{2}^\dagger V_{1}$ and $V_{1} V_{2}^\dagger = V_{2} V_{1}^\dagger$. This implies that $(V_{i}^\dagger V_{j})_{ij}$ and $(V_{j} V_{i}^\dagger)_{ij}$ are not jointly linearly independent. Again by Corollary \ref{c1} $\varrho$ is not extremal in $\mathcal{C}\left(\frac{1}{2} \openone, \frac{1}{2} \openone \right)$. A contradiction.

We summarize our results in the following proposition.
\begin{prop} \label{p1} In dimension $2 \times 2$ the extremal states in $\mathcal{C}\left(\frac{1}{2} \openone, \frac{1}{2} \openone \right)$ are precisely the projectors onto the subspaces spanned by maximally entangled pure wavefunctions. \end{prop} 
Proposition \ref{p1} has previously been found, using different methods, by Parthasarathy in \cite{Parthasarathy03}.

\subsection{A three dimensional example}
From the preceding example it is clear that also in higher dimensions all projectors onto the subspaces spanned by maximally entangled wavefunctions are extremal elements in $\mathcal{C}\left(\frac{1}{d} \openone, \frac{1}{d} \openone \right)$.
However, in the present section we show that the extension of Proposition \ref{p1} to higher dimensions does not hold. In other words the the set of extremal states in $\mathcal{C}\left(\frac{1}{d} \openone, \frac{1}{d} \openone \right)$ is not exhausted by the projectors onto maximally entangled pure states. Here we use our characterization of extremal states in $\mathcal{C}\left(\frac{1}{d} \openone, \frac{1}{d} \openone \right)$ to construct an explicit counterexample in dimension $3 \times 3$. 

Denote by $\{ \vert i \rangle \}_{i=1}^3$ the canonical real orthonormal basis of $\mathbb{C}^3$. Define the following operators
\addtocounter{equation}{1} \alpheqn
\begin{eqnarray} \label{wies}
V_{1} & = & \frac{1}{\sqrt{2}} \left( \vert 1 \rangle \langle 1 \vert + \vert 2 \rangle \langle 3 \vert \right) \\ \label{wies2}
V_{2} & = & \frac{1}{\sqrt{2}} \left( \vert 2 \rangle \langle 2 \vert + \vert 3 \rangle \langle 1 \vert \right) \\ \label{wies3}
V_{3} & = & \frac{1}{\sqrt{2}} \left( \vert 3 \rangle \langle 3 \vert + \vert 1 \rangle \langle 2 \vert \right). 
\end{eqnarray} \reseteqn
By explicit calculation one checks that $\sum_{j=1}^3 V_{j}^\dagger V_{j} = \sum_{j}^3 V_{j} V_{j}^\dagger = \openone$.
Moreover, \addtocounter{equation}{1} \alpheqn \begin{eqnarray} V_{1}^\dagger V_{2} & = & V_{3} V_{1}^\dagger = \frac{1}{2} \vert 3 \rangle \langle 2 \vert, \\
V_{1}^\dagger V_{3} & = & V_{3} V_{2}^\dagger = \frac{1}{2} \vert 1 \rangle \langle 2 \vert, \\
V_{2}^\dagger V_{3} & = & V_{1} V_{2}^\dagger = \frac{1}{2} \vert 1 \rangle \langle 3 \vert, \\
V_{2}^\dagger V_{1} & = & V_{1} V_{3}^\dagger = \frac{1}{2} \vert 2 \rangle \langle 3 \vert, \\
V_{3}^\dagger V_{1} & = & V_{2} V_{3}^\dagger = \frac{1}{2} \vert 2 \rangle \langle 1 \vert, \\
V_{3}^\dagger V_{2} & = & V_{2} V_{1}^\dagger = \frac{1}{2} \vert 3 \rangle \langle 1 \vert.
\end{eqnarray} \reseteqn
Hence $\{V_{i}^\dagger V_{j} \}_{ij}$ and $\{V_{j} V_{i}^\dagger \}_{ij}$ are both linearly independent and thus by Lemma \ref{l2} jointly linearly independent. By Corollary \ref{c1} the state
\begin{equation} \varrho := \frac{1}{3} \sum_{ijk=1}^3 V_{j}^\dagger \vert i \rangle \langle k \vert V_{j} \otimes \vert i \rangle \langle k \vert \end{equation} is extremal in $\mathcal{C}\left(\frac{1}{3} \openone, \frac{1}{3} \openone \right)$.
An explicit calculation gives the following matrix representation of $\varrho$ in the canonical product basis in lexicographic order
\begin{equation} \varrho = \frac{1}{6}  \left( \begin{array}{ccccccccc}
1 & 0 & 0 & 0 & 0 & 0 & 0 & 1 & 0 \\
0 & 0 & 0 & 0 & 0 & 0 & 0 & 0 & 0 \\
0 & 0 & 1 & 0 & 1 & 0 & 0 & 0 & 0 \\
0 & 0 & 0 & 1 & 0 & 0 & 0 & 0 & 1 \\
0 & 0 & 1 & 0 & 1 & 0 & 0 & 0 & 0 \\
0 & 0 & 0 & 0 & 0 & 0 & 0 & 0 & 0 \\
0 & 0 & 0 & 0 & 0 & 0 & 0 & 0 & 0 \\
1 & 0 & 0 & 0 & 0 & 0 & 0 & 1 & 0 \\
0 & 0 & 0 & 1 & 0 & 0 & 0 & 0 & 1 
\end{array} \right).
\end{equation}
This state is entangled but not maximally entangled and an extremal element of $\mathcal{C}\left(\frac{1}{3} \openone, \frac{1}{3} \openone \right)$.
\subsection{Higher dimensions}
It is possible to construct counterexamples to Proposition \ref{p1} also in higher dimensions. For instance consider ${\mathbb{C}}^4 \otimes {\mathbb{C}}^4$. We denote the canonical basis of ${\mathbb{C}}^4$ as usual by $\{ \vert 1 \rangle, \vert 2 \rangle, \vert 3 \rangle, \vert 4 \rangle \}$. 
The three dimensional example above can be  generalized to dimension $4 \times 4$ by letting 
\begin{eqnarray*} V_{1} & = & \frac{1}{\sqrt{3}} \left( \vert 1 \rangle \langle 1 \vert + \vert 2 \rangle \langle 4 \vert + \vert 3 \rangle \langle 2 \vert \right), \\
V_{2} & = & \frac{1}{\sqrt{3}} \left( \vert 2 \rangle \langle 2 \vert + \vert 3 \rangle \langle 1 \vert + \vert 4 \rangle \langle 3 \vert \right), \\
V_{3} & = & \frac{1}{\sqrt{3}} \left( \vert 3 \rangle \langle 3 \vert + \vert 4 \rangle \langle 2 \vert + \vert 1 \rangle \langle 4 \vert \right), \\
V_{4} & = & \frac{1}{\sqrt{3}} \left( \vert 4 \rangle \langle 4 \vert + \vert 1 \rangle \langle 3 \vert + \vert 2 \rangle \langle 1 \vert \right).
\end{eqnarray*} 
It is straightforward to show that both $(V_i^\dagger V_{j})_{ij}$ and $(V_{j} V_{i}^\dagger)_{ij}$ are linearly independent families. Thus an analysis similar to the one given above shows that \begin{equation} \varrho := \frac{1}{4} \sum_{ijk=1}^4 V_{j}^\dagger \vert i \rangle \langle k \vert V_{j} \otimes \vert i \rangle \langle k \vert \end{equation} is extremal in $\mathcal{C}\left(\frac{1}{4} \openone, \frac{1}{4} \openone \right)$ but is not a maximally entangled pure state.
It is easy to construct similar counterexamples also in higher dimensions. It seems therefore likely that there are counterexamples to Proposition \ref{p1} in all dimensions greater than 2. 
\acknowledgments
I am grateful to K.R.~Parthasarathy for explaining his work during the quantum information workshop at Pavia. 
Funding by the EC project ATESIT (contract IST-2000-29681) is acknowledged.

\end{document}